\documentclass[fleqn,usenatbib]{mnras}

\usepackage{txfonts}

\usepackage[T1]{fontenc}

\DeclareRobustCommand{\VAN}[3]{#2}
\let\VANthebibliography\thebibliography
\def\thebibliography{\DeclareRobustCommand{\VAN}[3]{##3}\VANthebibliography}

\usepackage{graphicx}	
\usepackage{color}
\usepackage{ulem}
\usepackage{upgreek}
\usepackage{longtable}

\title[VLBI astrometry on AR~Sco]{VLBI astrometry on the white dwarf pulsar AR~Scorpii}

\author[Jiang et al.]{Pengfei Jiang,$^{1,3}$ 
Lang Cui,$^{1,2}$\thanks{E-mail: \href{cuilang:cuilang@xao.ac.cn}{cuilang@xao.ac.cn}}
Jun Yang,$^{4}$
Bo Zhang,$^{5}$
Shuangjing Xu,$^{6,5}$
Fengchun Shu,$^{5}$
\newauthor Wu Jiang,$^{5}$
Wen Chen,$^{7,3}$
Guanghui Li,$^{1}$
Bo Xia,$^{5}$
Stuart Weston,$^{8}$
Sergei Gulyaev,$^{8}$
\newauthor Hongmin Cao,$^{9}$
Xiang Liu$^{1,2}$
and Tao An$^{5,1}$
\\
$^{1}$Xinjiang Astronomical Observatory, Chinese Academy of Sciences, 150 Science 1-Street, 830011 Urumqi, P. R. China\\
$^{2}$Key Laboratory of Radio Astronomy, Chinese Academy of Sciences, 150 Science 1-Street, 830011 Urumqi, P. R. China\\
$^{3}$School of Astronomy and Space Science, University of Chinese Academy of Sciences, 100049 Beijing, P. R. China\\
$^{4}$Department of Space, Earth and Environment, Chalmers University of Technology, Onsala Space Observatory, SE-439 92 Onsala, Sweden\\
$^{5}$Shanghai Astronomical Observatory, Chinese Academy of Sciences, 200030 Shanghai, P. R. China\\
$^{6}$Korea Astronomy and Space Science Institute, 776 Daedeokdae-ro, Yuseong-gu, Daejeon 34055, Republic of Korea\\
$^{7}$Yunnan Observatories, Chinese Academy of Sciences, 650216 Kunming, Yunnan, P.R. China\\
$^{8}$Institute for Radio Astronomy and Space Research, Auckland University of Technology, Private Bag 92006, Auckland 1142, New Zealand\\
$^{9}$School of Electronic and Electrical Engineering, Shangqiu Normal University, Wenhua Road 298, Shangqiu, Henan 476000, P. R. China
}

\date{Accepted 2023 January 25. Received 2023 January 24; in original form 2022 May 3}

\pubyear{2023}

\begin{document}
\label{firstpage}
\pagerange{\pageref{firstpage}--\pageref{lastpage}}
\maketitle

\begin{abstract}
AR Scorpii (AR Sco), the only-known radio-pulsing white dwarf binary, shows unusual pulsating emission at the radio, infrared, optical and ultraviolet bands. To determine its astrometric parameters at the radio band independently, we conducted multi-epoch Very Long Baseline Interferometry (VLBI) phase-referencing observations with the European VLBI Network (EVN) at 5\,GHz and the Chinese VLBI Network (CVN) plus the Warkworth 30-metre telescope (New Zealand) at 8.6\,GHz. By using the differential VLBI astrometry, we provide high-precision astrometric measurements on the parallax ($\pi=8.52_{-0.07}^{+0.04}$\,mas), and proper motion ($\mu_{\alpha}=9.48_{-0.07}^{+0.04}$\,mas\,yr$^{-1}$, $\mu_{\delta}=-51.32_{-0.38}^{+0.22}$\,mas\,yr$^{-1}$). The new VLBI results agree with the optical \textit{Gaia} astrometry. Our kinematic analysis reveals that the Galactic space velocities of AR Sco are quite consistent with that of both intermediate polars (IPs) and polars. Combined with the previous tightest VLBI constraint on the size, our parallax distance suggests that the radio emission of AR~Sco should be located within the light cylinder of its white dwarf.
\end{abstract}

\begin{keywords}
white dwarfs -- pulsars: individual: AR Sco -- parallaxes -- techniques: high angular resolution -- radio continuum: stars
\end{keywords}



\section{Introduction}

AR Scorpii (AR Sco) is a white dwarf (WD)/M-type star binary with an orbital period of $\sim3.56$\,h and a WD spin period of $\sim1.95$\,min \citep{Marsh2016}. Unusual pulsations in short timescales of a period of $\sim1.97$\,min are detected at multiple bands (ultraviolet, optical, infrared and radio). Its non-synchronous spin and orbital periods imply a connection with a class of stars known as intermediate polars \citep[IPs, e.g.][]{Patterson1994}, whereas its weak X-ray radiation distinguishes itself from usual IPs \citep{Takata2018}. The strong radio emission and broadband spectral energy distribution of AR Sco are similar to those of a special IP system AE Aquarii \citep{Bookbinder1987,Oruru2012}. However, the radio pulsation properties of AR~Sco were unique among WD systems.

The exact emission mechanism of AR~Sco is still unclear and a large number of models have been proposed to interpret its unique observed behaviour. \citet{Marsh2016} proposed scenarios where the pulsed emission came from collimated fast particle outflows or the direct interaction of the magnetosphere of the WD with the M companion. \citet{Geng2016} suggested that the WD's rotational axis was nearly perpendicular to its magnetic axis, and the magnetic interaction occurred when the WD's open field line beams swept the secondary's wind. \citet{Katz2017} also proposed a misaligned-spin model and supposed a precessing spin axis to explain the displacement of the optical maximum from conjunction of this system, but suggested that the magnetic interaction occurs within the MD's atmosphere. \citet{Garnavich2019} investigated the possible existence of magnetic loops or prominence of the M star, the magnetic interaction with these may contribute to the system's emission. The energy that comes from AR~Sco may be generated through magneto-hydrodynamic interactions \citep{Buckley2017} or fast magnetic reconnection events \citep{Garnavich2019}, and \citet{Bednarek2018} discussed a hadronic model.

Accurate astrometry on AR Sco will allow us to constrain its physical properties (e.g. velocity, luminosity, emission region size). It helps to validate the theoretical models, to trail the evolutionary history, and to explore the possibility of this system being a future source of detectable gravitational wave emission \citep{Franzon&Schramm2017}. For AR~Sco, \citet{Marsh2016} gave a rough estimate of the distance of $d = 116 \pm 16$\,pc, based on the spectra and photometry of the M star in optical and infrared. By combining with the first \textit{Gaia} Data Release \citep{Gaia2016b}, the fifth US Naval Observatory CCD Astrograph Catalog \citep{Zacharias2017} obtained a proper motion estimate of $\mu_{\alpha}=4.0\pm2.2$\,mas\,yr$^{-1}$, $\mu_{\delta}=-50.8\pm2.0$\,mas\,yr$^{-1}$ in optical.
Based on optical astrometric measurements, the \textit{Gaia} Early Data Release 3 \citep[\textit{Gaia} EDR3,][]{GaiaEDR3} provided a parallax estimate of $\pi=8.544\pm0.038$\,mas and proper motion estimates of $\mu_{\alpha}=9.690\pm0.047$\,mas\,yr$^{-1}$, $\mu_{\delta}=-51.489\pm0.038$\,mas\,yr$^{-1}$ for AR Sco. Our goal was to determine the astrometric parameters of AR~Sco with the technique of Very Long Baseline Interferometry (VLBI) at the radio band. As the only technique that can derive high precision positions of target sources comparable to \textit{Gaia}, VLBI could provide new and independent astrometric results to validate the \textit{Gaia} results.

In this paper, we present new astrometric measurements for AR Sco from the European VLBI Network (EVN) and the Chinese VLBI Network (CVN) plus Warkworth 30-m radio telescope observations over a period of $\sim1.5$ years. We describe the observations and data reduction in Section \ref{sec:section2}. The high precision astrometric results are presented in Section \ref{sec:section3}. We compare our VLBI results with those from \textit{Gaia}'s, and analyse the kinematics and physical parameters of AR~Sco in Section \ref{sec:section4}. Finally, we summarise the study in Section \ref{sec:section5}.

\section{Observations and data reduction}
\label{sec:section2} 

\subsection{EVN observations and data reduction}
\label{sec:section2.1}

Our EVN observations of AR~Sco at 5\,GHz were conducted in e-VLBI mode \citep{Szomoru2008} at five epochs between 2017 February and 2018 January. Table~\ref{tab:tab1} lists the observing dates, the participating stations and the time durations. The raw data were at a data rate of 2048\,Mbps (dual polarisation, 8$\times$32\,MHz bandwidth per polarisation, two-bit quantisation) and correlated in real-time (e-EVN mode) by the SFXC software correlator \citep{Keimpema2015} at JIVE (Joint Institute for VLBI ERIC, the Netherlands).

\begin{table*}
    \centering
    \caption{Summary of VLBI observations of AR Sco.}
    \label{tab:tab1}
    \begin{tabular}{ccclcc}
        \hline
        Project & Date         & Freq. & Participating Stations$^{a}$               & Duration  & Detection\\
        Code    &              & (GHz) &                                            & (h)       & \\
        \hline
        EL058A  & 2017 Feb 15 & 5.0   &\texttt{Ef, Jb, Mc, Nt, O8, Tr, Ys, Wb, Hh} & 4.4       & No  \\[2pt]
        EL058B  & 2017 Apr 12 & 5.0   &\texttt{Jb, Mc, Nt, O8, Tr, Ys, Wb, Hh, Ib} & 4.0       & Yes \\[2pt]
        EL058C  & 2017 Jun 20 & 5.0   &\texttt{Ef, Jb, Mc, Nt, Tr, Ys, Wb, Hh, Ir} & 4.0       & Yes \\[2pt]
        EL058D  & 2017 Sep 19 & 5.0   &\texttt{Ef, Jb, Nt, O8, Tr, Ys, Wb, Hh, Ir} & 3.8       & Yes \\ [2pt]
        EL058E  & 2018 Jan 17 & 5.0   &\texttt{Ef, Jb, Mc, Nt, O8, Tr, Hh, Ir}     & 3.9       & No  \\[2pt]
        \hline
        CC001A & 2017 Sep 25 & 8.6   &\texttt{Sh, T6, Km, Ur, Wa}                 & 6.5        & No  \\[2pt]
        CC001B & 2017 Dec 19 & 8.6   &\texttt{Sh, T6, Km, Ur, Wa}                 & 6.2        & Yes \\[2pt]
        CC001C & 2018 May 11 & 8.6   &\texttt{Sh, T6, Km, Ur, Wa}                 & 6.8        & Yes \\[2pt]
        CC001D & 2018 Jul 30 & 8.6   &\texttt{Sh, Km, Ur, Wa}                     & 6.9        & No  \\[2pt]
        CC001E & 2018 Sep 13 & 8.6   &\texttt{T6, Km, Ur}                         & 6.9        & Yes \\[2pt]
        \hline
    \end{tabular}\\
    \footnotesize{$^a$ \texttt{Ef}: Effelsberg (100\,m), \texttt{Jb}: Jodrell Bank MKII (38$\times$25\,m), \texttt{Mc}: Medicina (32\,m), \texttt{Nt}: Noto (32\,m), \texttt{O8}: Onsala-85 (25\,m), \texttt{Tr}: Torun (32\,m), \texttt{Ys}: Yebes (40\,m), \texttt{Wb}: Westerbork (25\,m), \texttt{Hh}: Hartebeesthoek (26\,m), \texttt{Ir}: Irbene (32\,m), \texttt{Ib}: Irbene (16\,m), \texttt{Sh}: Shanghai (25\,m), \texttt{T6}: Tianma (65\,m), \texttt{Km}: Kunming (40\,m), \texttt{Ur}: Urumqi (26\,m), \texttt{Wa}: Warkworth (30\,m).}\\
\end{table*}

The observations of AR~Sco were performed with the two phase-referencing calibrators: PKS~J1625$-$2527 \citep[J1625$-$2527, e.g.][]{Fey1996} and PMN~J1621$-$2241 \citep[J1621-2241,][]{Griffith1994}. The primary calibrator J1625$-$2527 located 2$\fdg$7 away from AR Sco was selected from \textit{Astrogeo Centre}\footnote{\url{http://astrogeo.org/calib/search.html}}. The secondary calibrator J1621$-$2241 was selected based on our pilot short VLBI observation (project code: RSC03) at 5\,GHz. It is 12\,arcmin away from AR Sco, and shows a point-like structure with a peak brightness of $\sim20$\,mJy\,beam$^{-1}$. The cycle time was about 6 minutes: $\sim1$\,min for the primary calibrator, $\sim4$\,min for the target, and $\sim1$\,min for the gap. The secondary calibrator was observed for one $\sim2$\,min scan per three cycles. This additional faint calibrator allows us to run a further iteration of the phase-referencing calibration to significantly improve the astrometric precision \citep[e.g.][]{Doi2006,Paragi2013}. Table~\ref{tab:tab2} lists the correlation phase centres for these three sources.

\begin{table}
    \centering
    \caption{Source correlation phase centres. $\theta_{\rm sep}$ give source separations from AR~Sco.}
    \label{tab:tab2}
    \begin{tabular}{llll}
        \hline
        Source           & $\alpha$ (J2000)                  & $\delta$ (J2000) & $\theta_{\rm sep}$\\
        \hline
        AR Sco$^{a}$            & $16^{\rm h}21^{\rm m}47\fs303367$ & $-22\degr53\arcmin11\farcs46486$ & ...\\[2pt]
        J1625$-$2527$^{b}$  & $16^{\rm h}25^{\rm m}46\fs891639$ & $-25\degr27\arcmin38\farcs32688$ & 2$\fdg$7 \\[2pt]
        J1621$-$2241 & $16^{\rm h}21^{\rm m}32\fs276276$ & $-22\degr41\arcmin01\farcs40904$ & 0$\fdg$2 \\ [2pt]
        \hline
    \end{tabular}\\
    \footnotesize{$^{a}$ The correlation phase centre is only for EL058B. $^{b}$ The position was taken from \url{http://astrogeo.org/vlbi/solutions/rfc_2015a/}.}\\
\end{table}

The data reduction was performed with the NRAO Astronomical Image Processing System \citep[\textsc{aips},][]{Greisen2003}. The baselines when one or both antennas were pointing below 10 degree elevation or the data were severely corrupted by radio frequency interference, weather, recording or instrumental problems were initially flagged. No correction was made to update the Earth Orientation Parameters or digital sampling bias corrections applied at correlation time by the SFXC correlator. The ionospheric delay was corrected using Jet Propulsion Laboratory Global Ionospheric Maps with the \textsc{aips} task \texttt{TECOR}. We conducted a priori amplitude calibration via standard gain curves and system temperature measurements of participating stations. In case of no available system temperature data, the priori amplitude calibration was conducted using the nominal system equivalent flux densities with the \textsc{aips} task \texttt{CLCOR}. The parallactic angle correction was applied by the task \texttt{CLCOR}. The instrumental phase errors across intermediate frequencies were removed through a manual phase calibration with the primary calibrator J1625$-$2527. Global fringe fitting and bandpass calibration were also applied with J1625$-$2527. The \textsc{aips} task \texttt{CALIB} was used to compute both amplitude and phase self-calibration corrections for J1625$-$2527, and the solutions were transferred to the secondary calibrator J1621$-$2241 and the target AR Sco. After these calibrations, the phases of both J1621$-$2241 and AR Sco were phase referenced to J1625$-$2527. Imaging and self-calibration for J1621$-$2241 data were performed in \textsc{difmap} \citep{Shepherd1994}. The clean maps for J1621$-$2241 without self-calibration were then loaded into \textsc{aips}, and the position and position uncertainties were derived with the task \texttt{JMFIT}. In our experiments, the secondary calibrator J1621$-$2241 was regarded as a stationary reference source. We set J1621$-$2241 at its correlation phase centre with a point model and used the \textsc{aips} task \texttt{CALIB} to derive phase corrections, and the corresponding solutions were applied to J1621$-$2241 and AR~Sco. This step could improve the final image fidelity of AR~Sco. And AR~Sco was phase referenced to J1621$-$2241. Finally, we imaged AR~Sco in \textsc{aips}.

\subsection{CVN plus \texttt{Wa} observations and data reduction}

In order to expand the time span of the observations and improve astrometric accuracy, we also conducted VLBI observations of AR~Sco at 8.6\,GHz with the CVN plus the Warkworth 30-m radio telescope \citep[\texttt{Wa,}][]{Woodburn2015}, located in New Zealand, under the program CC001 at five epochs between 2017 September and 2018 September. Adding \texttt{Wa} to the network allowed us to significantly boost the resolution in the north-south direction. The observing setup is also summarised in Table~\ref{tab:tab1}. The data were recorded in the disks at a data rate of 2048\,Mbps (16\,subbands, 32\,MHz filters, right-hand circular polarisation, two-bit quantisation). The correlation was executed with the DiFX software correlator \citep{Deller2011} at Shanghai Astronomical Observatory.

These CVN plus \texttt{Wa} observations at 8.6\,GHz followed the above EVN observing strategy. The data were also reduced and imaged in a very similar way. Moreover, the EOPs and digital sampler bias corrections were performed with the \textsc{aips} task \texttt{CLCOR} and \texttt{ACCOR}, respectively.

\section{Results}
\label{sec:section3}

\subsection{VLBI imaging results}
\begin{figure}
    \centering
    \includegraphics[width=0.8\columnwidth]{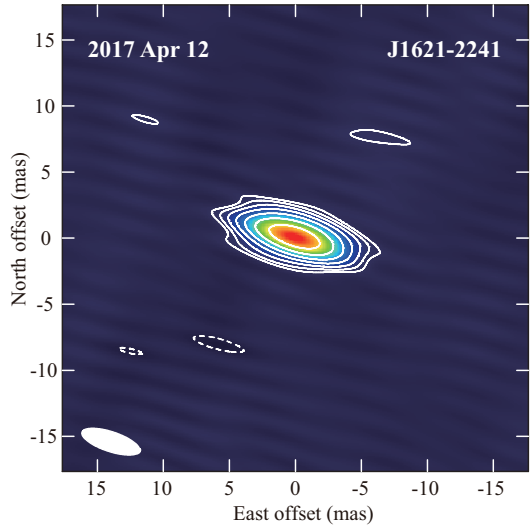}
    \caption{The EVN 5\,GHz image of the secondary phase-referencing calibrator J1621$-$2241. The image was obtained with natural weighting and self-calibration. The contours start from 0.3\,mJy\,beam$^{-1}$ and increase by a factor of two. The synthesised beam is plotted as the white ellipse in the bottom left corner and has a full width at half maximum (FWHM) of $4.6\times1.5$\,mas$^2$ at position angle PA = 71$\fdg$8.}
    \label{fig:fig1}
\end{figure}

The EVN image of the secondary calibrator J1621$-$2241 observed on 2017 April 12 is shown in Fig.~\ref{fig:fig1}. This image was made with natural weighting and has a noise level of 0.3\,mJy\,beam$^{-1}$. It shows a point-like source with a peak flux density of 36.4\,mJy\,beam$^{-1}$. Fitting a circular Gaussian model to the visibility data gave a flux density of $\sim41.3$\,mJy and a size of $\sim0.7$\,mas. For CVN plus \texttt{WA} observations at 8.6\,GHz, the calibrator also shows a point-like source with a mean total flux density of $\sim40.6$\,mJy and a mean size of $\sim1.2$\,mas. 

\begin{figure*}
    \includegraphics[width=\textwidth]{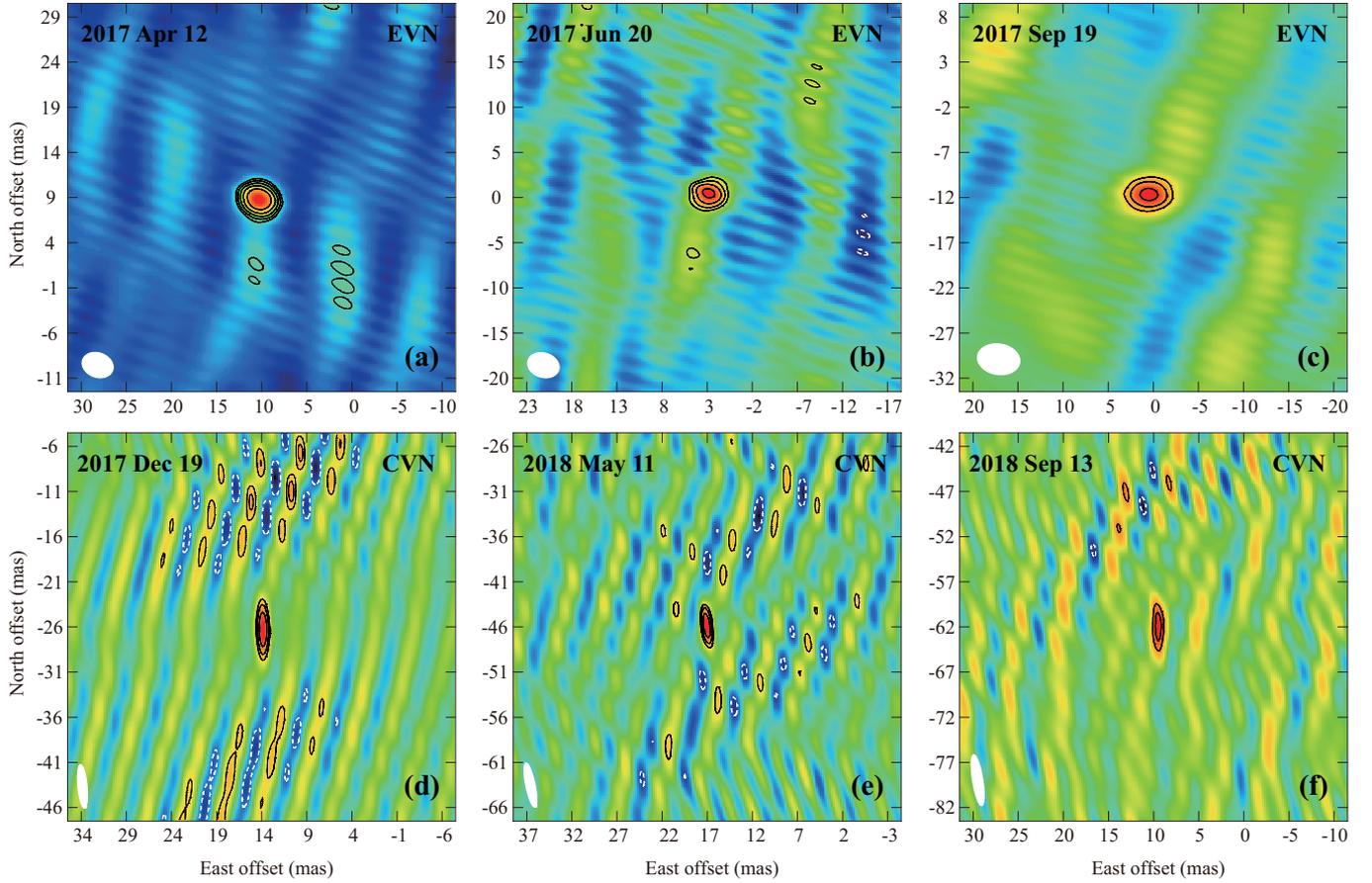}
    \caption{The VLBI \textsc{clean} maps of the white dwarf pulsar AR Sco at 5\,GHz (top) and 8.6\,GHz (bottom). The synthesised beam is shown in the bottom left corner of each panel. Contours start at three times the noise level of images and increase by factors of $\sqrt{2}$. The related information is listed in Table~\ref{tab:tab4}.}
    \label{fig:fig2}
\end{figure*}
 
Fig.~\ref{fig:fig2} shows the \textsc{clean} maps of the white dwarf pulsar AR~Sco at 5\,GHz in the top panels and at 8.6\,GHz in the bottom panels. All the images were produced with natural weighting. AR~Sco displays an unresolved structure at all epochs. Fitting a circular Gaussian model to the visibility data from the highest SNR epoch EL058B gave a size of $1.6\pm0.1$\,mas for AR~Sco. The related information is listed in Table~\ref{tab:tab3}. The images have relatively higher noise levels than the estimated thermal noise, especially for (c) and (f) listed in Table~\ref{tab:tab3}, probably due to a significant sensitivity loss of the 32\,MHz digital filters of the digital base band converter 2 system (DBBC2), which was also reported in other EVN observations at similar period \citep[e.g.][]{Yang2020}. To get more reasonable flux density estimates, we scaled the amplitudes by a factor of 1.7 and took a large fraction, 15 per cent, of the flux density as the uncertainty in Table~\ref{tab:tab3}. The factor was derived by comparing our flux densities obtained from CVN observations at 8.6\,GHz with that obtained from the Australian Long Baseline Array (LBA) observation at 8.4\,GHz \citep{Marcote2017}.

The target AR~Sco were successfully detected in six epochs. In the other four epochs (EL058A, EL058E, CC001A, CC001D), there were no useful astrometric results because of various issues (antenna failures, weak/no fringes for the calibrators, low observing elevations and incorrect coordinate).

\begin{table*}
    \caption{Summary of VLBI imaging results of AR Sco$^{a}$.}
    \label{tab:tab3}
    \begin{tabular}{ccccccccc}
        \hline
        Panel & MJD & FWHM  & PA        & $S_{\rm 5\,GHz}$ & $S_{\rm 8.6\,GHz}$ & SNR & $\Delta\alpha\cos\delta$ & $\Delta\delta$ \\
              &(day)&(mas)  & ($\degr$) & (mJy\,beam$^{-1}$)  & (mJy\,beam$^{-1}$)    &     & (mas)                    & (mas)           \\
        \hline
        (a)   &57855.105 & 3.5 $\times$ 2.7  & 70.1 & 7.0 $\pm$ 1.1 & ...             & 18.9 & $+10.290\pm0.085$ &  $+8.770\pm0.082$  \\[2pt]
        (b)   &57924.915 & 3.6 $\times$ 2.7  & 72.8 & 3.2 $\pm$ 0.6 & ...             & 8.9  & $+2.845\pm0.173$  &  $+0.332\pm0.157$  \\[2pt]
        (c)   &58015.664 & 4.7 $\times$ 3.4  & 81.0 & 3.2 $\pm$ 0.7 & ...             & 6.4  & $+0.453\pm0.351$  & $-11.573\pm0.263$  \\[2pt]
        (d)   &58107.134 & 4.8 $\times$ 1.0  & 5.4  & ...             & 4.1 $\pm$ 0.8 & 8.6  & $+13.865\pm0.076$ & $-26.277\pm0.261$  \\[2pt]
        (e)   &58249.725 & 5.0 $\times$ 1.0  & 11.6 & ...             & 6.5 $\pm$ 1.2 & 8.8  & $+17.022\pm0.074$ & $-45.928\pm0.186$  \\[2pt]
        (f)   &58374.439 & 5.7 $\times$ 1.2  & 9.1  & ...             & 9.4 $\pm$ 2.0 & 6.5  & $+9.523\pm0.097$  & $-62.039\pm0.327$  \\[2pt]
        \hline
    \end{tabular}\\
    \footnotesize{$^{a}$ Col. 1 - Panel code in Fig.~\ref{fig:fig2}, Col. 2 - Modified Julian Day (MJD), Col. 3 - size of the synthesized beam, Col. 4 - position angle, Col. 5 \& 6 - peak brightness at 5\,GHz or 8.6\,GHz, Col. 7 - signal to noise ratio (SNR) of images, and Col. 8 \& 9 - relative position offset.}\\
\end{table*}

\subsection{Astrometry}

The position of the extra-galactic reference source J1621$-$2241 was obtained from both the EVN observations and the CVN plus \texttt{Wa} observations, with respect to the primary calibrator J1625$-$2527. We found that the position dispersion in the EVN observations ($\sigma_{\alpha}=9.8$\,mas, $\sigma_{\delta}=8.9$\,mas) was larger than that in the CVN plus \texttt{Wa} observations ($\sigma_{\alpha}=1.0$\,mas, $\sigma_{\delta}=1.8$\,mas). For the EVN observations, the relatively high latitude of the array caused the difficulty of observing such low declination source, the majority of antennas were pointing at relatively low elevation ($\sim$ 15\degr) during observations. The large angular distance between the primary calibrator and the secondary could lead to lobe ambiguities for these low elevation observations. The limited on-source time ($\sim$ 20\,min) for the secondary calibrator could aggravate this condition. These terms cause the large position dispersion for J1621$-$2241 in the EVN observations. For the CVN plus \texttt{Wa} observation, the relative lower latitude array and longer  on-source time ($\sim$ 1\,h) for J1621$-$2241 alleviate this problem. Moreover, the lobe ambiguities will be eliminated for the target when referencing the target to the secondary calibrator, as the secondary calibrator is only 12\,arcmin away from the target.
 
A weighted average method was used to derive the position of J1621$-$2241 with the combination of EVN observations and CVN plus \texttt{Wa} observations. First, the arithmetic mean position was calculated from each of the two observational arrays separately. Next, we derived the weighted mean position from the two arithmetic mean positions. The weights of the mean positions from two arrays were set as inversely proportional to the square of their dispersion. As the used calibrator J1625$-$2527 is an International Celestial Reference Frame (ICRF) defining source, its position was corrected by its ICRF3\footnote{\url{https://hpiers.obspm.fr/icrs-pc/newwww/index.php}} \citep{Charlot2020} S/X-band coordinate of $\alpha_{J2000} =16^{\rm h}25^{\rm m}46\fs8916429\pm0.03$~mas, $\delta_{J2000} =-25\degr27\arcmin38\farcs326873\pm0.03$~mas. Finally, we provided the position of J1621$-$2241 of $\alpha_{J2000}=16^{\rm h}21^{\rm m}32\fs27524\pm1.0$\,mas, $\delta_{J2000}=-22\degr41\arcmin01\farcs3982\pm1.7$\,mas, with the uncertainty given by the combination of the uncertainty in the weighted mean position of J1621$-$2241 and the uncertainty in the ICRF3 position of J1625$-$2527.

With respect to J1621$-$2241, we measured the astrometric parameters at the epoch J2018 for AR~Sco. The \textsc{aips} task \texttt{JMFIT} was used to derive the radio centroid of AR~Sco at each epoch. The final position offsets and the formal position uncertainties are listed in Table~\ref{tab:tab3}. These formal position uncertainties are the combinations of the statistical error extracted in the image by the \textsc{aips} task \texttt{JMFIT} and the astrometric error caused by the phase-referencing technology. For the latter, we derived an estimate of 0.07\,mas from \citet{Pradel2006} based on the position offset between the target and the secondary calibrator. For the astrometric fit, the effect caused by the potential core shift in the secondary calibrator was considered. No obvious jet structure was detected for the secondary calibrator in our observations. The typical frequency-dependent core shift of 0.11\,mas in each direction between 5\,GHz and $\sim$8.6\,GHz \citep{Sokolovsky2011} was added as a term of systemic error to the position uncertainties which were derived from EVN observations for AR~Sco.

\begin{figure*}
    \includegraphics[width=\textwidth]{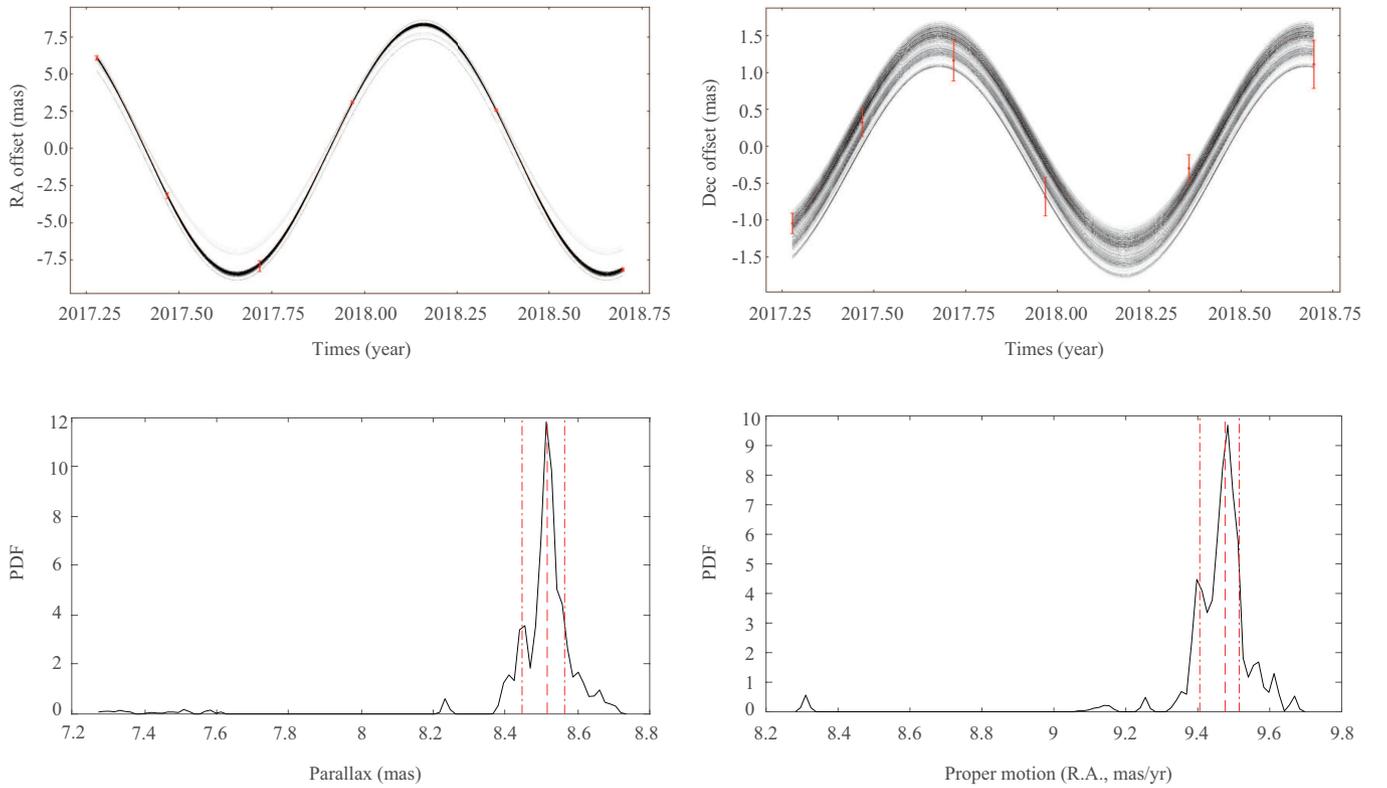}
    \caption{Top: illustration of the bootstrap astrometric fit for AR~Sco, showing position offset in right ascension (left) and declination (right) with the proper motion removed. Each of the 10\,000 trial fits is plotted in light grey line. Bottom: probability distribution functions (PDF) of parallax (left) and proper motion in right ascension (right). The median values and the 68 per cent confidence intervals are indicated with red vertical dashed lines and red vertical dot-dashed lines, respectively.}
    \label{fig:fig3}
\end{figure*}

A bootstrap approach \citep[e.g.][]{Efron1991,Deller2019} for astrometric fit was performed on the position offsets and their uncertainties for AR Sco. We conducted 10\,000 times random sampling with replacement from the available six data points. A least-squares fit was performed for each sampled data set. At last, a statistical distribution was obtained for each parameter. Fig.~\ref{fig:fig3} shows the bootstrap astrometric fit and bootstrap histograms for AR~Sco. According to the median value and the statistical 68 per cent confidence interval, we derived a parallax of $\pi = 8.52_{-0.07}^{+0.04}$\,mas. The other astrometric parameters were also derived in the same way and the results are listed in Table~\ref{tab:tab4}. Note that we used both the \textit{Gaia} EDR3 position and our VLBI position of J1621$-$2241 to calculate the position of AR~Sco, so each result is with respect to the celestial reference frame of \textit{Gaia} \citep[\textit{Gaia}-CRF,][]{Gaia2022} and ICRF3, respectively. The position uncertainties of J1621$-$2241 were taken into account for estimating the position uncertainties of AR Sco. For the \textit{Gaia} EDR3 position of J1621$-$2241, a median radio–optical positional offset of 0.5\,mas from \citet{Gaia2022} was added to the position uncertainties.

\begin{figure}
    \includegraphics[width=\columnwidth]{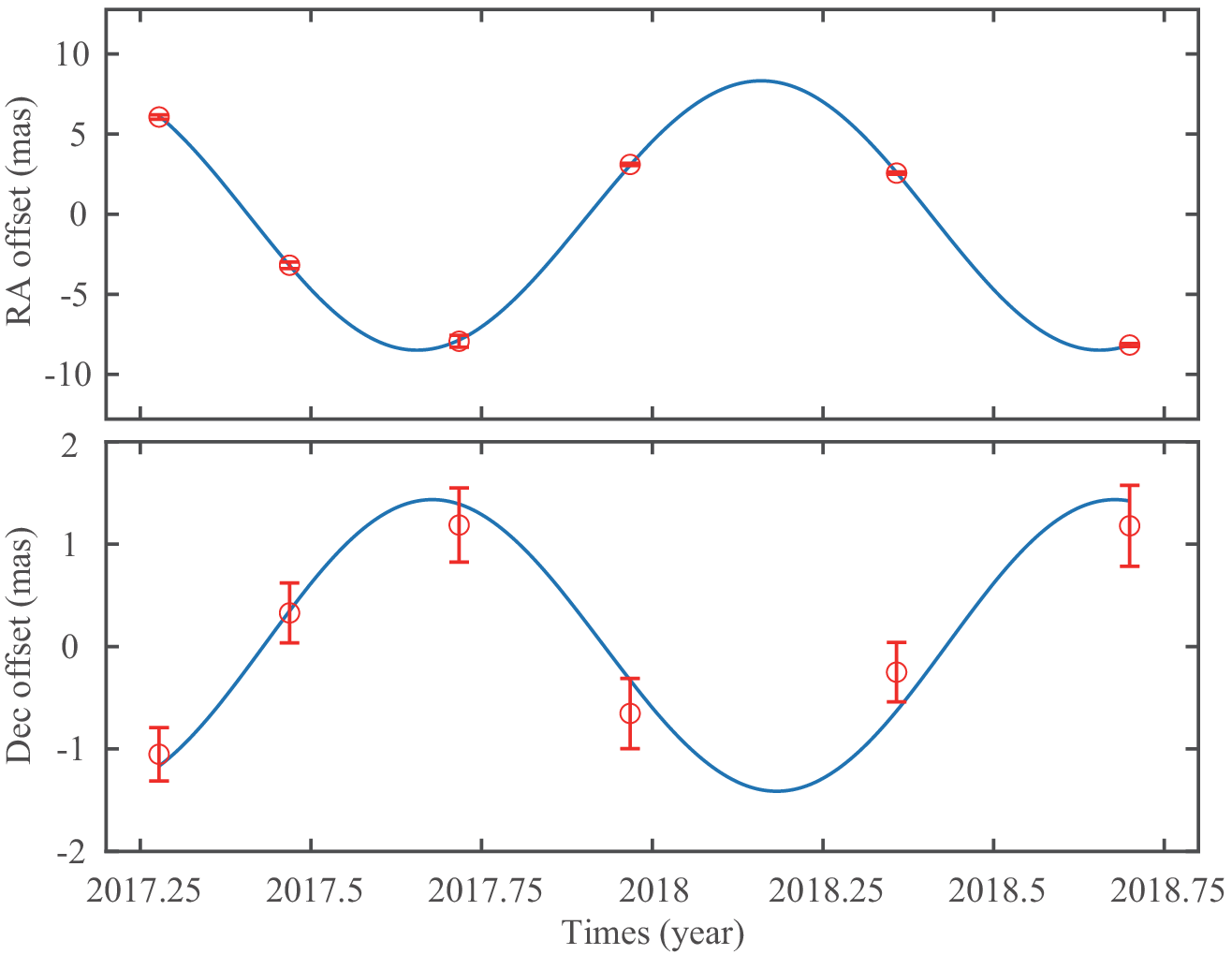}
    \caption{The least-squares astrometric fit for AR Sco, showing position offset in right ascension (upper) and declination (bottom) with the proper motion removed.}
    \label{fig:fig4}
\end{figure}

The reduced chi-squared of the least-squares fit was performed to verify the bootstrap astrometric fit. The illustration of the final astrometric fit is shown in Fig.~\ref{fig:fig4}. We obtained a parallax estimate of $\pi=8.53\pm0.07$\,mas and proper motion estimates of $\mu_{\alpha}=9.47\pm0.09$\,mas\,yr$^{-1}$, $\mu_{\delta}=-51.37\pm0.19$\,mas\,yr$^{-1}$ for AR~Sco. All astrometric parameters derived by the reduced chi-squared method are well consistent with those extracted from the bootstrap approach. Therefore, the reduced chi-squared astrometric fit validates the bootstrap results. And we report the bootstrap results as our final results.

\begin{table*}
    \caption{Fitted astrometric parameters for AR Sco at epoch J2018.}
    \label{tab:tab4}
    \begin{tabular}{lccc}
        \hline
        Parameter & This work & This work & \textit{Gaia} Collaboration \\
                  & (w.r.t ICRF3) & (w.r.t \textit{Gaia}-CRF) & et al. 2021\\
        \hline
        $\alpha_{0}$ (\rm{h\,m\,s})   & $16\,21\,47.29576_{-0.00007}^{+0.00007}$ & $16\,21\,47.295729_{-0.000070}^{+0.000070}$ & $16\,21\,47.295820_{-0.0000072}^{+0.0000072}$   \\[2pt]
        $\delta_{0}$ ($\degr\,\arcmin\,\arcmin \arcmin$)   & $-22\,53\,11.3164_{-0.0018}^{+0.0018}$ & $-22\,53\,11.31303_{-0.00075}^{+0.00072}$ & $-22\,53\,11.31249_{-0.000080}^{+0.000080}$   \\[2pt]
        $\mu_{\alpha}$ (mas\,yr$^{-1}$)   & $9.48_{-0.07}^{+0.04}$ & $9.48_{-0.07}^{+0.04}$ & $9.690_{-0.047}^{+0.047}$   \\[2pt]
        $\mu_{\delta}$ (mas\,yr$^{-1}$)   & $-51.32_{-0.38}^{+0.22}$ & $-51.32_{-0.38}^{+0.22}$ & $-51.489_{-0.038}^{+0.038}$ \\[2pt]
        $\pi$ (mas)$^{a}$  & $8.52_{-0.07}^{+0.04}$ & $8.52_{-0.07}^{+0.04}$ & $8.586_{-0.040}^{+0.040}$   \\[2pt]
        \hline
    \end{tabular}\\
    \footnotesize{$^a$ A zero-point correction has been applied to the \textit{Gaia} parallax.}\\
\end{table*}

\section{Discussion}
\label{sec:section4}

\subsection{Comparison with results of \textit{Gaia} EDR3}
\label{subsec:subsection4.1}

We obtained independent astrometric measurements for AR~Sco at the radio band in Section \ref{sec:section3}. In optical, \textit{Gaia} EDR3 provided astrometric results for AR~Sco as listed in Table~\ref{tab:tab4}. The \textit{Gaia} position was calculated to epoch J2018 and a direct comparison with our VLBI results can be performed. The position parameters for the VLBI results were derived in both ICRF3 and \textit{Gaia}-CRF. The differences between the VLBI positions and the \textit{Gaia}'s are smaller than or close to 1-$\sigma$ uncertainties in \textit{Gaia}-CRF. In ICRF3, the difference in position in R.A. is smaller than 1-$\sigma$ uncertainty and the difference in position in decl. is quite close to 2-$\sigma$ uncertainty. We note that the uncertainties in the position of AR Sco are dominated by those of its reference source J1621$-$2241. Future astrometric VLBI observations for J1621$-$2241 will give improved estimates in the position of AR~Sco. At present, there is no obvious offset between the optical and radio positions of AR~Sco.

For the {Gaia} parallax, a zero-point correction of $-0.043$\,mas from \citet{Lindegren2021} and an inflated error bar of 0.040\,mas from \citet{Badry2021} has been applied to the raw \textit{Gaia} parallax $\pi=8.544\pm0.038$\,mas. We found that the VLBI parallax was consistent with the raw \textit{Gaia} parallax within 1-$\sigma$. But the difference between the VLBI parallax and the zero-point corrected \textit{Gaia} parallax slightly exceeds 1-$\sigma$. The global parallax bias for {Gaia} EDR3 measured from a large number of quasars is 17\,$\upmu$as \citep{Lindegren2021}. Compared with the global parallax bias, the value of the zero-point correction of $-0.043$\,mas for AR~Sco is relatively large. AR~Sco is situated at small Galactic latitude ($\sim19\degr$) could explain it, in which a small number of quasars identified and a possible contamination of the quasars sample caused by Galactic stars leads to the difficulty of estimates of zero-point corrections. In addition, the zero-point estimates are well populated by the
quasars at magnitude G$\geq$16. Beyond that region, the uncertainties of the zero-point corrections will be greater \citep{Lindegren2021}. The apparent magnitude of AR Sco in the \textit{Gaia} G band is 14.99, beyond that region. VLBI astrometric observations of radio stars can validate the quality of \textit{Gaia}-CRF and help to determine the parallax zero-point of \textit{Gaia} catalogue \citep[e.g.][]{Bobylev2019,Xu2019,Lindegren2020}. Our high-precision VLBI results for AR~Sco can be used as independent measurements for the aims.

In a word, the VLBI results are consistent with the \textit{Gaia} result with no obvious offset. Here we derived astrometric parameters for AR Sco with the combination of VLBI and \textit{Gaia} result. The weighted average method was used to derive new astrometric parameter. The weights of astrometric parameters were set to be inversely proportional to the square of their uncertainties during the calculation. We obtain the parallax ($\pi=8.57\pm0.03$\,mas), and proper motion ($\mu_{\alpha}=9.62\pm0.04$\,mas\,yr$^{-1}$, $\mu_{\delta}=-51.49\pm0.04$\,mas\,yr$^{-1}$), for AR~Sco. 

\subsection{Kinematics}
\label{subsec:subsection4.2}

We computed the Galactic space velocities $U$, $V$ and $W$ for AR~Sco in a right-handed system following \citet{Johnson1987J} where $U$, $V$ and $W$ were measured positive in the directions of the Galactic centre, the Galactic rotation, and the North Galactic Pole, respectively. An assumption of zero radial velocity for the target AR~Sco was adopted to calculate the Galactic space velocities \citep[e.g.][]{Sion2014}. A motion of the Sun of ($U_{\bigodot}$, $V_{\bigodot}$, $W_{\bigodot}$) = (10.1, 13.6, 7.0)\,km\,s$^{-1}$ \citep{Bobylev2021} relative to the local standard of rest  was adopted for the solar motion correction. By using the astrometric results obtained in \ref{subsec:subsection4.1}, we derived Galactic space velocities of ($U$, $V$, $W$) = (15.4, $-$4.8, $-$14.7)\,km\,s$^{-1}$, corresponding to a total velocity of $T$ = 21.8\,km\,s$^{-1}$, for AR~Sco.

For AR~Sco, its non-synchronous spin and orbital periods imply a connection with IPs. \citet{Takata2018} categorised AR~Sco as an unusual IP directly. However, the lack of Doppler-broadened line emission is indicative of the absence of accretion disks, and the weak X-ray radiation implies that the majority of its luminosity is not attributable to accretion. In contrast, accretion is the main power source for most IPs \citep[e.g.][]{Buckley2000}. A more reliable hypothesis is that AR Sco represents an evolutionary stage of IPs \citep{Marsh2016}. In this sense, AR~Sco could be a progenitor star of IPs \citep{Meintjes2017} or a transitional star lying between IPs and traditional polars \citep{Katz2017}. Moreover, \citet{Schreiber2021} presented evolutionary models for the magnetic WDs in close binary stars and suggested that AR Sco is a progenitor star of polars. AR Sco represents a short spin down phase for strongly magnetic WDs in binaries in their evolutionary sequence. The stage occurs at the beginning of synchronization of the spin and orbital period for strongly magnetic WD binaries and finally polars will be formed.

\begin{figure}
    \includegraphics[width=\columnwidth]{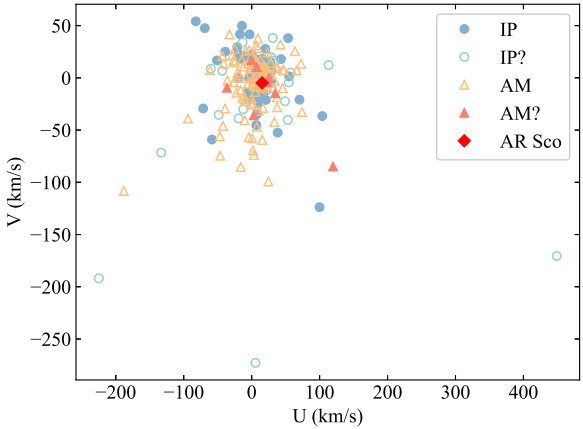}
    \caption{$U-V$ space velocity diagram for AR Sco and the sample of IPs and polors.}
    \label{fig:fig5}
\end{figure}

Here we investigate the kinematics of AR~Sco relative to a sample of 82 IPs and 107 polars. The kinematics and statistics of IPs and polars can be seen in appendix~\ref{appendixA}. Statistics show that IPs have significantly lower velocity dispersion than IP candidates. Polars have slightly lower velocity dispersion than polar candidates. The kinematic properties of IPs and polars are highly similar to each other, as both IPs and polars are sub-type stars of magnetic cataclysmic variables. In Fig.~\ref{fig:fig5}, we plotted $U-V$ space velocity diagram for the sample and AR Sco. Being an unique source, AR Sco, however, is not an outlier with respect to both IPs and polars as shown in Fig.~\ref{fig:fig5}. Its Galactic space velocities are quite consistent with that of both IPs and polars. This is reasonable under the assumptions that AR Sco now goes through a short-lived spin-down evolutionary stage of IPs or polars.

\subsection{Radio emitting region}

The parallax derived from the combinations of VLBI and \textit{Gaia} results sets AR Sco at a distance of $116.7\pm0.4$\,pc. With a source size of 0.17\,mas of AR Sco obtained with the LBA at 8.4\,GHz reported by \citet{Marcote2017}, the distance establishes that the radio emitting region size is $\lesssim 3\times10^{11}$\,cm. The size is only about half the light cylinder radius ($R_{\rm LC}\sim6\times10^{11}$\,cm) of the WD. This implies that the main radio emission region is situated inside the light cylinder of the WD. The finding is in agreement with what \citet{Singh2020} derived from $\gamma$-ray data and favours the models in which the main radio emission comes from a region near the WD, or the M star as this companion is situated well inside the magnetosphere of the WD \citep[e.g.][]{Geng2016,Katz2017}.

\subsection{Orbital reflex motion}

For AR~Sco, the mass of the WD and M star were derived as $M_{\rm WD}\approx0.8M_{\bigodot}$, $M_{\rm M}\approx0.3M_{\bigodot}$, respectively \citep{Marsh2016}. With the measured orbital period $P=0.14853528(8)$\,d and circular orbit, the orbital radius of the WD is $R_{\rm WD}\sim2\times10^{10}$\,cm, and the orbital radius of the secondary star is $R_{\rm M}\sim6\times10^{10}$\,cm. Based on the distance of AR Sco, an amplitude of possible reflex motion of $\sim40$\,$\upmu$as could be expected under the assumption that the non-thermal emission originates from the MD's atmosphere \citep[e.g.][]{Katz2017}. For those scenarios in which the radio emission region close to the MD surface \citep[e.g.][]{Plessis2022}, a smaller amplitude of $\sim10$\,$\upmu$as may occur. In any case, the amplitude is below the measurement accuracy for our observations. The effect caused by the orbital motion is negligible based on the positions of AR~Sco were derived with integral time close to or longer than the orbital period in our measurements. Additionally, the calculations imply that the possible detection of the orbital reflex motion for the non-thermal emission could help to constrain the emitting region location for AR~Sco. The microarcsecond VLBI astrometry with high sensitivity will make an effort in the future. For instance, the $\sim10$\,$\upmu$as goal of astrometry using VLBI with the Square Kilometer Array was suggested to be feasible by using the MultiView technique \citep{Rioja2017}.

\subsection{Revisiting other physical parameters}

Based on the M star's spectra and brightness, \citet{Marsh2016} derived a distance estimate $d\approx116[M_{\rm M}/(0.3M_{\bigodot})]^{1/3}$\,pc for AR Sco, assuming that the M star was close to its Roche lobe and its mass $M_{\rm M}\approx0.3M_{\bigodot}$. Adopting the distance of 116\,pc to the spectral energy distribution of this system, they obtained a maximum luminosity of $L_{\rm max}\approx6.3\times10^{25}$\,W. They also measured the orbital period $P=0.14853528(8)$\,d and the radial velocity amplitude of the M star $K=295\pm4$\,km\,s$^{-1}$, and defined the mass function
\begin{equation}
    \frac{M_{\rm WD}^{3}\sin^{3}{i}}{(M_{\rm WD}+M_{\rm M})^{2}}=\frac{PK^{3}}{2\pi G}=(0.395\pm0.016)M_{\bigodot},
    \label{eq:eq1}
\end{equation}
where $i$ is the orbital inclination.

Our measurements obtained a more accurate distance for AR~Sco. A mass of $M_{\rm M}\approx0.31M_{\bigodot}$ for the M star was derived from the distance-mass relation $d\approx116[M_{\rm M}/(0.3M_{\bigodot})]^{1/3}$ with our distance. Following the assumption suggested by \citet{duPlessis2019} that the orbital inclination $i$ is approximately equal to the observer angle $\zeta=60_{\cdot}^{\circ}4_{-6_{\cdot}^{\circ}0}^{+5_{\cdot}^{\circ}3}$, we solve the mass function (\ref{eq:eq1}) with $M_{\rm M}\approx0.31M_{\bigodot}$, and derived $M_{\rm WD}\approx1.0_{-0.1}^{+0.2}M_{\bigodot}$ for the WD. However, the observer angle provided by \citet{duPlessis2019} was derived from polarization position angle data which were averaged over a large range of orbital phase. An updated work performed by \citet{Plessis2022} shows a variation of $\sim30\degr$ in $\zeta$ over the orbital period, the $\zeta$ is approximately in range from $55\degr$ to $85\degr$. Based on these results, we provide a more conservative estimate of $M_{\rm WD}\approx1.0 \pm 0.2 M_{\bigodot}$ for the WD. Our distance also resets the maximum luminosity of $L_{\rm max}\approx6.4\times10^{25}$\,W for AR~Sco.

Our parallax supports the distance estimated by \citet{Marsh2016}, and the recalculation shows that no substantial correction is needed to these physical parameters of AR Sco.

\section{Summary}
\label{sec:section5}

With the EVN observations at 5\,GHz and the CVN plus \texttt{Wa} observations at 8.6\,GHz, we clearly detected the compact radio emission of the only-known white dwarf pulsar AR~Sco in the epochs between 2017 February and 2018 September. Using a nearby calibrator, 12\,arcmin apart from AR~Sco, we provide high-precision measurements on the parallax ($\pi=8.52_{-0.07}^{+0.04}$\,mas), and proper motion ($\mu_{\alpha}=9.48_{-0.07}^{+0.04}$\,mas\,yr$^{-1}$, $\mu_{\delta}=-51.32_{-0.38}^{+0.22}$\,mas\,yr$^{-1}$), for AR~Sco. The independent measurements are consistent with the \textit{Gaia} EDR3 reports. Kinematic analysis shows that the Galactic space velocities of AR Sco are quite consistent with that of intermediate polars and polars. Together with the early tightest VLBI constraint on the size, our parallax distance suggests that the radio emission of AR~Sco would be located within the light cylinder of its white dwarf. Furthermore, we revisit the mass and luminosity of AR Sco with the parallax distance and confirm the previous estimates reported by \citet{Marsh2016}.

\section*{Acknowledgements}

We thank Sandor Frey, Kazuhiro Hada and the anonymous referee for helpful discussions and constructive suggestions, which resulted in an overall improvement of the paper.
This work is supported by the CAS `Light of West China' Program (grant No. 2021-XBQNXZ-005) and the National Natural Science Foundation of China (grant No. U2031212, 11673051 and U1831136). L. C. is also thankful for the support of the Youth Innovation Promotion Association of the CAS (No. 2017084).
The European VLBI Network is a joint facility of independent European, African, Asian, and North American radio astronomy institutes. Scientific results from data presented in this publication are derived from the following EVN project codes: RSC03 and EL058. 
e-VLBI research infrastructure in Europe is supported by the European Union’s Seventh Framework Programme (FP7/2007-2013) under grant agreement number RI-261525 NEXPReS. 
We are thankful for the observation time of the Chinese VLBI Network and the Warkworth 30m radio telescope operated by the Institute for Radio Astronomy and Space Research, Auckland University of Technology.
The computing cluster of Shanghai VLBI correlator supported by the Special Fund for Astronomy from the Ministry of Finance in China is acknowledged.
This work has made use of data from the European Space Agency (ESA) mission {\it Gaia} (\url{https://www.cosmos.esa.int/gaia}), processed by the {\it Gaia} Data Processing and Analysis Consortium (DPAC, \url{https://www.cosmos.esa.int/web/gaia/dpac/consortium}). Funding for the DPAC has been provided by national institutions, in particular the institutions participating in the {\it Gaia} Multilateral Agreement. 
This research has made use of the SIMBAD database, operated at CDS, Strasbourg, France.

\section*{Data Availability}

The correlated data of the experiments RSC03 and EL058 are available in the EVN data archive (\url{http://archive.jive.nl/scripts/portal.php}). The correlated data of the experiments CC001 can be requested from the corresponding author. The \textit{Gaia} EDR3 data underlying this article are available in the \textit{Gaia} Archive (\url{https://gea.esac.esa.int/archive/}). The ICRF3 catalogue is available at the website of the International Earth Rotation and Reference Systems Service (\url{https://hpiers.obspm.fr/icrs-pc/newwww/index.php}).



\bibliographystyle{mnras}
\bibliography{example} 




\appendix

\section{Kinematics and statistics of IPs and Polars}
\label{appendixA}

Table~\ref{tab:tabA1} presents a sample of IPs and polars with Galactic space motion results. We assembled this sample by starting with a traversal search on the \textit{Catalogue of Cataclysmic Binaries, Low-Mass X-Ray Binaries and Related Objects} \citep[Edition 7.24,][]{Ritter2003}. We obtained a sample of 280 stars for IPs and polars and found that 189 stars already had \textit{Gaia} EDR3  results (the position, the parallax and the proper motion) be identified in the SIMBAD astronomical database \citep{Wenger2000}. The sample of 189 stars consists of 82 IPs and 107 polars. The Galactic space velocity calculations for this sample followed the same method described in Subsection~\ref{subsec:subsection4.2}. The resultant space motion parameters are listed in Table~\ref{tab:tabA1} and the corresponding statistical information is listed in Table~\ref{tab:tabA2}. 

\onecolumn
\begin{longtable}{lcrrrrllcrrrr}
    \caption{Kinematics for the sample of IPs and polars.}
    \label{tab:tabA1}\\
    \hline 
    \multicolumn{1}{l}{Name} & \multicolumn{1}{c}{Type$^{a}$} & \multicolumn{1}{c}{$U$} & \multicolumn{1}{c}{$V$} & \multicolumn{1}{c}{$W$} & \multicolumn{1}{c}{$T$} & \qquad\qquad\qquad &\multicolumn{1}{l}{Name} & \multicolumn{1}{c}{Type$^{a}$} & \multicolumn{1}{c}{$U$} & \multicolumn{1}{c}{$V$} & \multicolumn{1}{c}{$W$} & \multicolumn{1}{c}{$T$}\\
    \hline
    \endfirsthead
    
    \multicolumn{4}{c}{{\tablename\ \thetable{} $--$ continued from previous page}} \\
    \hline
    \multicolumn{1}{l}{Name} & \multicolumn{1}{c}{Type$^{a}$} & \multicolumn{1}{c}{$U$} & \multicolumn{1}{c}{$V$} & \multicolumn{1}{c}{$W$} & \multicolumn{1}{c}{$T$} & \qquad\qquad\qquad &\multicolumn{1}{l}{Name} & \multicolumn{1}{c}{Type$^{a}$} & \multicolumn{1}{c}{$U$} & \multicolumn{1}{c}{$V$} & \multicolumn{1}{c}{$W$} & \multicolumn{1}{c}{$T$}\\
    \hline
    \endhead
    
    \hline \multicolumn{3}{l}{{Continued on next page}} \\
    \endfoot
    
    \multicolumn{13}{l}{
     $^{a}$ IP - intermediate polar, IP? - intermediate polar candidate. AM - polar, AM? - polar candidate. The types of sources were obtained}\\
    \multicolumn{13}{l}{
     from \citet{Ritter2003} if there are no additional references. (1)\citet{Worpel2020}, (2) \citet{Hilton2009},}\\
     \multicolumn{13}{l}{
      (3) \citet{Halpern2022}, (4) \citet{Sambruna1992}, (5) \citet{Kubiak1994}, (6) \citet{Yakin2013}, }\\
       \multicolumn{13}{l}{
      (7) \citet{Rosen1995}, (8) \citet{Papadaki2006}, (9) \citet{Zemko2018}. }
    \endlastfoot
    
GK Per	&	IP	&	11.4 	&	2.4 	&	-29.7 	&	31.9 	& &	J1740$-$2847	&	IP	&	10.0 	&	-0.4 	&	4.6 	&	11.0 	\\
V2731 Oph	&	IP	&	-3.0 	&	41.6 	&	21.5 	&	46.9 	& &	EX Hya	&	IP	&	-19.4 	&	-0.2 	&	13.9 	&	23.9 	\\
V1062 Tau	&	IP	&	13.8 	&	11.6 	&	-13.6 	&	22.5 	& &	DW Cnc	&	IP	&	-2.0 	&	18.7 	&	-14.5 	&	23.8 	\\
NY Lup	&	IP	&	-4.2 	&	-21.1 	&	-20.1 	&	29.4 	& &	HT Cam	&	IP	&	8.1 	&	9.6 	&	7.0 	&	14.3 	\\
V902 Mon	&	IP	&	19.8 	&	-1.3 	&	-10.3 	&	22.3 	& &	FS Aur	&	IP	&	10.6 	&	-23.7 	&	34.1 	&	42.8 	\\
J0838$-$4831	&	IP	&	-51.0 	&	16.7 	&	14.3 	&	55.6 	& &	V1025 Cen	&	IP	&	-71.4 	&	-29.3 	&	15.5 	&	78.8 	\\
V2069 Cyg	&	IP	&	26.8 	&	12.7 	&	6.6 	&	30.4 	& &	CC Scl	&	IP	&	70.3 	&	-20.9 	&	25.4 	&	77.6 	\\
J0457+4527	&	IP	&	11.0 	&	17.1 	&	-4.9 	&	20.9 	& &	J0503$-$2823	&	IP	&	-69.0 	&	47.6 	&	42.6 	&	94.0 	\\
J2133+5107	&	IP	&	27.5 	&	15.0 	&	10.3 	&	33.0 	& &	J0525+2413	&	IP	&	10.4 	&	-1.2 	&	7.9 	&	13.1 	\\
EI UMa	&	IP	&	-1.5 	&	-0.5 	&	-5.1 	&	5.3 	& &	J0614+1704	&	IP	&	17.1 	&	-15.2 	&	1.4 	&	22.9 	\\
J1509$-$6649	&	IP	&	-4.2 	&	0.1 	&	0.0 	&	4.2 	& &	J2015+3711	&	IP	&	55.1 	&	1.4 	&	9.2 	&	55.9 	\\
V3037 Oph	&	IP	&	9.8 	&	-10.9 	&	8.0 	&	16.7 	& &	AE Aqr	&	IP	&	-11.1 	&	20.7 	&	-14.7 	&	27.7 	\\
WX Pyx	&	IP	&	-19.7 	&	22.7 	&	-16.9 	&	34.5 	& &	DQ Her	&	IP	&	-17.8 	&	18.1 	&	14.6 	&	29.3 	\\
TV Col	&	IP	&	-39.0 	&	25.0 	&	36.3 	&	58.9 	& &	V349 Aqr	&	$^{1}$IP	&	99.9 	&	-123.8 	&	-81.7 	&	178.8 	\\
V709 Cas	&	IP	&	9.4 	&	12.5 	&	-2.8 	&	15.9 	& &	V598 Peg	&	$^{2}$IP	&	43.2 	&	17.9 	&	7.5 	&	47.4 	\\
PQ Gem	&	IP	&	-14.3 	&	49.8 	&	-9.7 	&	52.7 	& &	J1832$-$0840	&	$^{3}$IP	&	38.2 	&	-52.4 	&	-1.6 	&	64.9 	\\
HZ Pup	&	IP	&	-82.3 	&	54.1 	&	-3.3 	&	98.5 	& &	QR And	&	IP?	&	-224.9 	&	-192.0 	&	-105.9 	&	314.1 	\\
FO Aqr	&	IP	&	0.3 	&	22.2 	&	7.8 	&	23.5 	& &	J0939$-$3226	&	IP?	&	-60.2 	&	8.7 	&	-45.3 	&	75.8 	\\
MU Cam	&	IP	&	16.0 	&	18.2 	&	10.8 	&	26.5 	& &	V2275 Cyg	&	IP?	&	113.3 	&	12.2 	&	13.2 	&	114.7 	\\
HY Leo	&	IP	&	20.8 	&	-21.2 	&	-7.4 	&	30.6 	& &	V426 Oph	&	IP?	&	29.4 	&	-10.0 	&	-7.6 	&	31.9 	\\
V1323 Her	&	IP	&	103.9 	&	-36.6 	&	26.1 	&	113.2 	& &	V1039 Cen	&	IP?	&	-48.5 	&	-35.4 	&	9.3 	&	60.8 	\\
V418 Gem	&	IP	&	6.2 	&	11.3 	&	-9.8 	&	16.2 	& &	AH Eri	&	IP?	&	53.2 	&	-40.3 	&	-8.3 	&	67.3 	\\
V2306 Cyg	&	IP	&	29.3 	&	6.9 	&	-13.3 	&	32.9 	& &	AP Cru	&	IP?	&	-133.3 	&	-71.6 	&	3.7 	&	151.3 	\\
LS Peg	&	IP	&	1.0 	&	-1.1 	&	-22.3 	&	22.4 	& &	J1616$-$4958	&	IP?	&	6.5 	&	6.5 	&	4.3 	&	10.2 	\\
V405 Aur	&	IP	&	-6.1 	&	-14.0 	&	-14.0 	&	20.7 	& &	V4745 Sgr	&	IP?	&	5.5 	&	-272.9 	&	-48.0 	&	277.1 	\\
V1033 Cas	&	IP	&	53.5 	&	38.0 	&	-4.7 	&	65.8 	& &	GI Mon	&	IP?	&	-13.3 	&	34.0 	&	-31.5 	&	48.3 	\\
J1719$-$4100	&	IP	&	7.0 	&	-1.1 	&	13.8 	&	15.5 	& &	CW Mon	&	IP?	&	17.8 	&	-0.7 	&	21.0 	&	27.5 	\\
DO Dra	&	IP	&	20.1 	&	27.7 	&	2.6 	&	34.3 	& &	J1446+0253	&	IP?	&	49.1 	&	-22.4 	&	-23.6 	&	58.8 	\\
J0153+7446	&	IP	&	10.5 	&	12.5 	&	12.2 	&	20.4 	& &	V2467 Cyg	&	IP?	&	57.5 	&	4.9 	&	18.2 	&	60.5 	\\
V842 Cen	&	IP	&	-58.5 	&	-59.0 	&	4.9 	&	83.3 	& &	LS Cam	&	IP?	&	31.0 	&	38.2 	&	9.0 	&	50.0 	\\
J1654$-$1916	&	IP	&	7.7 	&	-34.8 	&	20.8 	&	41.2 	& &	V592 Cas	&	IP?	&	28.9 	&	19.8 	&	-22.7 	&	41.7 	\\
J1649$-$3307	&	IP	&	4.7 	&	-18.3 	&	3.7 	&	19.3 	& &	GZ Cnc	&	IP?	&	-12.4 	&	-30.0 	&	-55.9 	&	64.6 	\\
AO Psc	&	IP	&	10.4 	&	-23.1 	&	-18.3 	&	31.2 	& &	VZ Pyx	&	IP?	&	0.6 	&	12.0 	&	-12.1 	&	17.1 	\\
V647 Aur	&	IP	&	13.6 	&	-1.2 	&	23.5 	&	27.1 	& &	BZ UMa	&	IP?	&	19.6 	&	1.4 	&	23.4 	&	30.6 	\\
UU Col	&	IP	&	-21.3 	&	29.4 	&	13.9 	&	38.8 	& &	YY Sex	&	IP?	&	21.3 	&	-12.2 	&	-14.7 	&	28.6 	\\
J1926+1322	&	IP	&	28.5 	&	-2.8 	&	-1.9 	&	28.7 	& &	QZ Vir	&	IP?	&	-19.3 	&	-38.8 	&	-24.3 	&	49.7 	\\
J2014+1529	&	IP	&	21.8 	&	6.2 	&	8.6 	&	24.3 	& &	V533 Her	&	IP?	&	0.6 	&	17.7 	&	6.0 	&	18.7 	\\
V2400 Oph	&	IP	&	8.6 	&	17.3 	&	16.8 	&	25.6 	& &	KO Vel	&	$^{4, 5}$IP?	&	-43.4 	&	6.7 	&	10.8 	&	45.2 	\\
V1223 Sgr	&	IP	&	6.8 	&	-45.2 	&	-25.7 	&	52.5 	& &	J2216+4646	&	$^{6}$IP?	&	33.6 	&	16.4 	&	5.3 	&	37.7 	\\
BG CMi	&	IP	&	-17.2 	&	41.8 	&	-37.5 	&	58.7 	& &	V795 Her	&	$^{7, 8}$IP?	&	54.6 	&	-4.3 	&	-7.0 	&	55.2 	\\
V515 And	&	IP	&	16.7 	&	11.1 	&	-12.2 	&	23.5 	& &	V2491 Cyg	&	$^{9}$IP?	&	449.0 	&	-170.5 	&	5.9 	&	480.3 	\\
V479 And	&	AM	&	63.7 	&	25.5 	&	-16.6 	&	70.6 	& & 	EK UMa	&	AM	&	-46.3 	&	-74.6 	&	-1.2 	&	87.8 	\\
V1309 Ori	&	AM	&	16.6 	&	-9.6 	&	12.3 	&	22.8 	& & 	ST LMi	&	AM	&	17.7 	&	-4.1 	&	5.7 	&	19.0 	\\
AI Tri	&	AM	&	22.7 	&	-8.1 	&	-33.1 	&	40.9 	& & 	BL Hyi	&	AM	&	-18.6 	&	2.4 	&	4.8 	&	19.4 	\\
J0649-0737	&	AM	&	20.4 	&	1.3 	&	5.5 	&	21.1 	& & 	MR Ser	&	AM	&	-28.6 	&	21.6 	&	33.1 	&	48.8 	\\
MQ Dra	&	AM	&	1.8 	&	-5.8 	&	25.6 	&	26.3 	& & 	FR Lyn	&	AM	&	1.8 	&	-69.5 	&	-8.5 	&	70.1 	\\
J2048+0050	&	AM	&	5.8 	&	17.6 	&	7.3 	&	19.9 	& & 	V884 Her	&	AM	&	-8.5 	&	26.0 	&	20.9 	&	34.4 	\\
V1043 Cen	&	AM	&	3.1 	&	-8.1 	&	-15.3 	&	17.6 	& & 	V2301 Oph	&	AM	&	-3.6 	&	26.6 	&	21.7 	&	34.5 	\\
0922+1333	&	AM	&	14.1 	&	-2.0 	&	-0.7 	&	14.2 	& & 	CD Ind	&	AM	&	-16.4 	&	-12.0 	&	-9.4 	&	22.4 	\\
VY For	&	AM	&	0.0 	&	-19.9 	&	28.2 	&	34.5 	& & 	J1002$-$1925	&	AM	&	-25.3 	&	14.9 	&	-5.6 	&	29.9 	\\
J0227+1306	&	AM	&	36.2 	&	9.4 	&	-20.0 	&	42.4 	& & 	EP Dra	&	AM	&	-61.2 	&	7.3 	&	-7.3 	&	62.0 	\\
QQ Vul	&	AM	&	27.8 	&	3.3 	&	-0.3 	&	28.0 	& & 	J0953+1458	&	AM	&	42.5 	&	31.9 	&	41.3 	&	67.3 	\\
J0749-0549	&	AM	&	20.7 	&	-3.3 	&	-17.9 	&	27.5 	& & 	J0706+0324	&	AM	&	37.4 	&	-28.2 	&	22.3 	&	51.9 	\\
V358 Aqr	&	AM	&	5.1 	&	-13.6 	&	-9.7 	&	17.4 	& & 	V834 Cen	&	AM	&	-22.5 	&	-14.3 	&	24.3 	&	36.0 	\\
J1007-2017	&	AM	&	-12.1 	&	10.7 	&	-6.4 	&	17.4 	& & 	VV Pup	&	AM	&	45.6 	&	-9.8 	&	-7.8 	&	47.3 	\\
V388 Peg	&	AM	&	37.2 	&	-18.7 	&	-21.1 	&	46.6 	& & 	EG Lyn	&	AM	&	0.7 	&	-57.8 	&	12.2 	&	59.1 	\\
J1422-0221	&	AM	&	19.5 	&	8.6 	&	-0.8 	&	21.3 	& & 	J1344+2044	&	AM	&	24.6 	&	-99.2 	&	7.4 	&	102.5 	\\
V1432 Aql	&	AM	&	16.9 	&	10.2 	&	22.7 	&	30.1 	& & 	V393 Pav	&	AM	&	-9.9 	&	-20.6 	&	-4.5 	&	23.3 	\\
BY Cam	&	AM	&	9.1 	&	37.6 	&	-36.5 	&	53.2 	& & 	HS Cam	&	AM	&	2.3 	&	-3.9 	&	10.9 	&	11.8 	\\
V1500 Cyg	&	AM	&	72.6 	&	13.4 	&	13.5 	&	75.1 	& & 	LW Cam	&	AM	&	18.5 	&	11.5 	&	24.9 	&	33.1 	\\
J0733+2619	&	AM	&	11.6 	&	1.4 	&	3.5 	&	12.2 	& & 	BS Tri	&	AM	&	11.0 	&	-8.1 	&	-15.6 	&	20.7 	\\
J0837+3830	&	AM	&	-13.1 	&	-19.6 	&	-27.0 	&	35.9 	& & 	EQ Cet	&	AM	&	-4.2 	&	-56.9 	&	11.7 	&	58.2 	\\
V519 Ser	&	AM	&	22.1 	&	9.3 	&	-6.7 	&	24.9 	& & 	J1944$-$4202	&	AM	&	6.6 	&	31.0 	&	-1.3 	&	31.7 	\\
J1453-5521	&	AM	&	-47.0 	&	-56.0 	&	-15.8 	&	74.8 	& & 	J1312+1736	&	AM	&	23.4 	&	24.2 	&	5.8 	&	34.2 	\\
CW Hyi	&	AM	&	-39.2 	&	-29.5 	&	29.7 	&	57.3 	& & 	J1321+5609	&	AM	&	46.9 	&	7.6 	&	18.7 	&	51.1 	\\
J2319+2615	&	AM	&	14.7 	&	5.2 	&	-7.3 	&	17.2 	& & 	EU UMa	&	AM	&	-5.7 	&	-5.6 	&	1.7 	&	8.2 	\\
HY Eri	&	AM	&	33.4 	&	-42.2 	&	8.7 	&	54.5 	& & 	V347 Pav	&	AM	&	2.9 	&	18.2 	&	-10.6 	&	21.2 	\\
WX LMi	&	AM	&	20.5 	&	-3.3 	&	13.1 	&	24.5 	& & 	J0257+3337	&	AM	&	7.8 	&	2.0 	&	-1.7 	&	8.2 	\\
EU Lyn	&	AM	&	6.0 	&	13.3 	&	-1.0 	&	14.6 	& & 	J0502+1624	&	AM	&	9.8 	&	-31.7 	&	22.9 	&	40.3 	\\
V349 Pav	&	AM	&	8.8 	&	28.4 	&	-6.2 	&	30.4 	& & 	DP Leo	&	AM	&	-26.4 	&	-4.2 	&	-9.0 	&	28.2 	\\
PZ Vir	&	AM	&	-35.7 	&	-2.5 	&	19.4 	&	40.7 	& & 	CP Tuc	&	AM	&	0.8 	&	-24.3 	&	17.9 	&	30.2 	\\
J0524+4244	&	AM	&	-0.5 	&	-25.8 	&	-4.2 	&	26.2 	& & 	J1514+0744	&	AM	&	3.3 	&	-73.8 	&	24.5 	&	77.8 	\\
AP CrB	&	AM	&	-14.4 	&	17.4 	&	19.7 	&	30.0 	& & 	V379 Vir	&	AM	&	16.7 	&	-28.9 	&	-15.1 	&	36.6 	\\
V654 Aur	&	AM	&	13.1 	&	10.5 	&	13.8 	&	21.7 	& & 	IW Eri	&	AM	&	66.2 	&	7.2 	&	-41.2 	&	78.4 	\\
J0859+0536	&	AM	&	33.6 	&	-2.9 	&	16.8 	&	37.7 	& & 	J1250+1549	&	AM	&	-10.4 	&	-40.6 	&	0.0 	&	41.9 	\\
QS Tel	&	AM	&	-7.9 	&	0.1 	&	-24.5 	&	25.8 	& & 	J0425$-$5714	&	AM	&	-41.2 	&	13.5 	&	10.0 	&	44.5 	\\
V516 Pup	&	AM	&	-43.1 	&	20.8 	&	4.4 	&	48.0 	& & 	GQ Mus	&	AM	&	-93.8 	&	-39.0 	&	-1.2 	&	101.6 	\\
V381 Vel	&	AM	&	37.1 	&	11.1 	&	-11.9 	&	40.6 	& & 	J0921+2038	&	AM	&	-24.9 	&	27.9 	&	-19.5 	&	42.2 	\\
V1189 Her	&	AM	&	-8.5 	&	-22.4 	&	50.4 	&	55.8 	& & 	BM CrB	&	AM	&	18.6 	&	3.8 	&	9.8 	&	21.3 	\\
UW Pic	&	AM	&	-5.5 	&	13.5 	&	14.5 	&	20.5 	& & 	IL Leo	&	AM	&	15.7 	&	-59.5 	&	-18.3 	&	64.2 	\\
J1333+1437	&	AM	&	-188.3 	&	-108.3 	&	50.6 	&	223.0 	& & 	EF Eri	&	AM	&	-15.9 	&	-85.4 	&	55.4 	&	103.1 	\\
HU Leo	&	AM	&	-32.6 	&	41.5 	&	-7.4 	&	53.3 	& & 	J0154$-$5947	&	AM	&	28.0 	&	14.6 	&	10.4 	&	33.2 	\\
J2218+1925	&	AM	&	28.1 	&	3.3 	&	-5.0 	&	28.8 	& & 	J0528+2838	&	AM	&	7.6 	&	-46.9 	&	17.8 	&	50.7 	\\
MT Dra	&	AM	&	-26.7 	&	16.9 	&	6.5 	&	32.3 	& & 	GG Leo	&	AM	&	-20.6 	&	5.2 	&	-15.2 	&	26.2 	\\
UZ For	&	AM	&	1.6 	&	-0.7 	&	18.7 	&	18.8 	& & 	EV UMa	&	AM	&	-11.5 	&	-16.9 	&	15.7 	&	25.8 	\\
J2218+1925	&	AM	&	5.6 	&	-5.5 	&	14.5 	&	16.5 	& & 	V4738 Sgr	&	AM	&	12.9 	&	16.9 	&	11.2 	&	24.0 	\\
EU Cnc	&	AM	&	-2.3 	&	8.6 	&	-13.8 	&	16.4 	& & 	J2340+7642	&	AM?	&	-36.4 	&	-9.5 	&	-1.2 	&	37.6 	\\
HU Aqr	&	AM	&	74.1 	&	-32.8 	&	26.9 	&	85.4 	& & 	J0759+1914 	&	AM?	&	25.3 	&	-0.4 	&	27.2 	&	37.1 	\\
J1743$-$0429 	&	AM	&	19.6 	&	7.5 	&	-22.3 	&	30.7 	& & 	J0935+1619	&	AM?	&	3.1 	&	-35.8 	&	-29.4 	&	46.5 	\\
J0328+0522	&	AM	&	27.0 	&	-4.8 	&	-13.8 	&	30.7 	& & 	CP Pup	&	AM?	&	-0.4 	&	16.8 	&	6.7 	&	18.1 	\\
V2951 Oph	&	AM	&	24.9 	&	-1.0 	&	-11.7 	&	27.6 	& & 	J0311$-$3152	&	AM?	&	34.8 	&	-14.8 	&	10.6 	&	39.2 	\\
V808 Aur	&	AM	&	0.3 	&	10.0 	&	-16.2 	&	19.0 	& & 	J1955+0045	&	AM?	&	22.6 	&	-4.1 	&	-1.9 	&	23.1 	\\
V1237 Her	&	AM	&	46.1 	&	7.2 	&	-5.9 	&	47.1 	& & 	PT Per	&	AM?	&	7.4 	&	10.2 	&	0.8 	&	12.7 	\\
AR UMa	&	AM	&	-22.9 	&	0.2 	&	-6.7 	&	23.8 	& & 	J0354$-$1652	&	AM?	&	119.8 	&	-84.9 	&	-38.3 	&	151.7 	\\
AN UMa	&	AM	&	-42.2 	&	-46.0 	&	-12.1 	&	63.6 	& & 												\\
        \hline
        \end{longtable}
\twocolumn

\begin{table}
    \centering
    \caption{Kinematical Statistics of IPs and polars.}
    \label{tab:tabA2}
    \begin{tabular}{lccrr}
        \hline
        Type & Number  & Component & Average & Dispersion \\
        \hline
  IP   & 57 & $U$         & 6.4    & 34.4 \\[2pt]
       &    & $V$         & 1.7    & 29.9 \\[2pt]
       &    & $W$         & 0.9    & 19.9 \\[2pt]
       &    & $T$         & 39.7   & 30.3 \\[2pt]
  IP?  & 25 & $U$         & 16.7   & 112.0 \\[2pt]
       &    & $V$         & -28.9  & 74.9 \\[2pt]
       &    & $W$         & -11.1   & 29.4 \\[2pt]
       &    & $T$         & 88.7   & 110.0 \\[2pt]
  IP+IP? & 82 & $U$       & 9.5    & 67.5 \\[2pt]
       &    & $V$         & -7.6   & 49.8 \\[2pt]
       &    & $W$         & -2.8   & 23.7 \\[2pt]
       &    & $T$         & 54.6   & 68.8 \\[2pt]
  AM   & 99 & $U$         & 2.6    & 34.2 \\[2pt]
       &    & $V$         & -7.4    & 29.5 \\[2pt]
       &    & $W$         & 3.0    & 18.5 \\[2pt]
       &    & $T$         & 40.4   & 28.4 \\[2pt]
  AM?  & 8 & $U$         & 22.0   & 45.1 \\[2pt]
       &    & $V$         & -15.3  & 32.4 \\[2pt]
       &    & $W$         & -3.2   & 21.2 \\[2pt]
       &    & $T$         & 45.8   & 44.4 \\[2pt]
  AM+AM? & 107 & $U$       & 4.0    & 35.2 \\[2pt]
       &    & $V$         & -8.0   & 29.7 \\[2pt]
       &    & $W$         & 2.5   & 18.7 \\[2pt]
       &    & $T$         & 40.8   & 29.6 \\[2pt]     
        \hline
    \end{tabular}\\
\end{table}


\bsp	
\label{lastpage}
\end{document}